# Partial ablation of Ti/Al nano-layer thin film by single femtosecond laser pulse

B. Gaković[*,1], G. D. Tsibidis[2,3], E. Skoulas[2], S. Petrović[1], B. Vasić[4] and E. Stratakis[2,3]

[1] Institute of Nuclear Sciences "Vinča", University of Belgrade, P.O. Box 522, 11001 Belgrade, Serbia

[2] Institute of Electronic Structure and Laser (IESL), Foundation for Research and Technology (FORTH), N. Plastira 100, Vassilika Vouton, 70013, Heraklion, Crete, Greece

[3] Materials Science and Technology Department, University of Crete, 71003 Heraklion, Greece

[4] Institute of Physics, University of Belgrade, Pregrevica 118, 11080 Beograd, Serbia

**Abstract**

The effects of ultra-short laser pulses on reactive Ti/Al nano-layered thin film were investigated. The thin film composed of alternated titanium and aluminium nano-layers, was deposited by ion-sputtering. Single pulse irradiation was conducted in the air with focused and linearly polarized femtosecond laser beam - of 1026 nm wavelength and pulse duration of 170 fs. Laser induced composition and morphological changes, using different microscopy techniques and energy dispersive X-ray spectroscopy, were investigated. Following results were obtained: (i) one step partial/selective ablation of upper Ti layer from nano-layer Ti/Al at low laser fluence and (ii) two step ablation or entire ablation of nano-layer Ti/Al at higher laser fluence. Single pulse selective ablation of the upper Ti layer was confirmed based on profiling (AFM) along the ablation steps and reduction of Ti concentration (EDX) in the ablated areas. Ablation threshold was estimated using well known procedure for ultra-short laser pulses - spot diameter square versus logarithm of pulse energy. To interpret the experimental observations, simulations have been performed to explore the thermal response of the multiple layered structure (Ti(5x(Al/Ti))) after irradiation with a characteristic value of single laser pulse of fluence $F = 320$ mJ/cm$^2$. The results are in agreement with the calculations.

# 1. Introduction

Processing of materials with femtosecond (fs) lasers, in general, has received considerable attention over the past decades due to its important applications in modern technology[1,2,3]. Applications require a comprehensive knowledge of the laser interaction with particular targets for enhanced controllability of the resulting modification. One of the potentials of using fs pulsed lasers is possibility for high-precision reproducible modification of thin films[4,5]. Partial ablation/exfoliation of the thin film, with little or no damage of its substrate beneath, is important for applications in damage-free machining process of the CIS-based photovoltaic devices with the selective laser scribing process by simultaneous ablation[6,7,8,9]. One of the most important effects of fs laser processing is that the formation of a heat affected zone (HAZ) is suppressed due to the extremely short pulse widths of several tens to several hundred fs[10]. Thus, high-precision, high-quality micro- and nano fabrication can be performed even for high thermal conductivity materials such as metals. Nano scale multilayered thin films represent an attractive area for study due to their unusual properties, such as enormous hardness or a typical phase constitution, which cannot be obtained in uniform bulk materials[11,12,13,14]. On the other hand, if the nano-layered thin film is composed of reactive metals its processing should be considered with special attention because of fact that constituents can exothermically react. Upon external energetic source these reactive multi layers can release stored chemical energy. This sudden exothermic reaction may become self-sustained and can propagate along the thin film surface alternating the huge area. More than 50 reactive multilayered thin films (RMTF) have been known up to now and their physic-chemical properties were investigated[15]. Laser beam, as an energetic source is nowadays used for starting intermixing[16,17] of reactants in the RMTF, but only few papers are dealing with ultra-short laser pulses effects on RMTF[18,19,20]. In case of application of laser beam for micro processing the extensive area of intermixing should be avoid, due to the creation of gradient structure composed of a composite with good mechanical performance on the substrate and alloy on the surface as a protective layer. Solution is usage of fs laser pulses at low fluence in which case only very small area/volume of RMTF is affected by irradiation. Taking into account that small number of reactive multilayered systems was investigated after fs laser irradiation studying of different metals combination in RMTF might be useful for wide range applications. In selective laser ablation of thin films, the precise control of the ablation in the vertical direction becomes critical in order to avoid the damage of the deeper layers and substrate, due to adheres the desired distribution of components in the depth. Also, with the selective laser ablation of multilayer structure can be achieved patterning of surface in means of different composition needed for flexible solar cells[7,8].

In our study, as a model system, the Ti/Al RMTF with its excellent mechanical and tribological properties[16,21] was chosen. The sample prepared for irradiation was the sandwich structure Ti(5x(Al/Ti) with entire thickness of about 200 nm. The first deposited layer was Ti due to less expressed surface oxidation in atmosphere compare with Al. We showed that single fs pulse irradiation with pulse energy/fluence of 0.5 µJ/0.32 Jcm$^{-2}$ was able to precisely ablate the first Ti layer from the nano-layer structure without producing significant modification of the layers beneath and non-irradiated area of the surface. We investigated how an increase of pulse energy affected the morphological and compositional changes of the ablated nano-layer thin film. The ablation threshold, one very important parameter for laser processing, was also computed. To interpret the experimental observations, simulations have been performed to explore the thermal response of the multiple layered structure (Ti(5x(Al/Ti))) after irradiation with a characteristic value of single laser pulse fluence F = 320 mJ/cm$^2$. The theoretical predictions are in a very good agreement with the
experimental results.

## 2. Experimental

### *2.1 Thin film deposition*

The experimental sample, the Ti(5x(Al/Ti)) multilayered structure, was prepared by a commercial Balzers Sputtron II vacuum system using 1.5 keV argon ions and 99.9% pure Al and Ti targets. The substrate used was n-type silicon (100) wafer (0.5 mm thick). It was cleaned by HF etch and dip in deionized water before mounting in the chamber. Prior to the thin film deposition, the substrate was additionally cleaned by back-sputtering. Deposition was performed in a single vacuum run without heating the substrate. The thickness of multilayered structure was about 200 nm. It consisted of five bilayer Al/Ti (thickness of individual Ti and Al layer was 17 nm) and covered with Ti top layer thickness of 27 nm. In the rest of this work, this multilayered thin film will be denoted as Ti(5x(Al/Ti))/Si.

### *2.2 Fs laser processing*

In the experiment we used an Yb:KGW laser. Irradiation was performed with focused, linearly p-polarized beams in air. Characteristics of the laser beam were 1026 nm central wavelength, 170 fs pulse duration and repetition rate of 1 kHz. The spatial distribution was Gaussian with $1/e^2$ beam radius of 34 μm. For purpose of precise selective ablation the beam was focused up to ∼ 10 μm. Irradiation was done with pulse energies ($E_p$) varied from 0.2 μJ to 10 μJ. The sample Ti(5x(Al/Ti))/Si was mounted on a motorized, computer controlling, X–Y-Z translation stage, normal to the incident laser beam. Irradiation was done with choosing appropriate translation velocity, to obtain single laser pulse effects on the surface with the varied pulse energy. For statistics, the irradiations were conducted by identical conditions in a row consisted of 5 spots. In each row, pulse energy $E_p$ was assumed constant as deviation was less than 1%.

### *2.3 Characterization*

Various analytical techniques were used to study morphology and composition of the as-deposited thin film and laser irradiated regions. Surface of the sample, before and after laser processing, was examined firstly by optical microscopy and then more detailed by scanning electron microscopy (SEM). In addition to optical microscopy and SEM, the morphology of the Ti(5x(Al/Ti)) surface was also examined by atomic force microscopy (AFM). Atomic force microscopy (AFM) imaging was done by NTEGRA Prima system operating in tapping mode and using NSG01 probes. All measurements were done at ambient conditions. An AFM quantitative analysis of irradiated areas was performed to measure depth of created shallow holes or the ablated areas steps. On the other hand, elemental-compositional characterization of the sample was performed by energy-dispersive X-ray spectroscopy (EDX).

## 3. Results and Discussion

Firstly, before irradiations, we performed examination of as deposited thin Ti(5x(Al/Ti)) film. By means, of optical microscopy and SEM analysis at low magnification, it was revealed that the deposited thin film exhibited mirror like surface. At higher magnification, SEM and AFM analysis showed the fine grainy structure of the thin film surface with a mean surface roughness ($R_{RMS}$) of approximately 4.8 nm.

In the case of ultrafast laser irradiation, physical processes depend on the sample characteristics and on the laser beam properties. It is well known that laser processing of material surface with ultra-short laser pulses[22] result in well-defined ablated areas. The heat-affected zone is not as huge as in the case of longer pulses and irradiation lefts behind a negligible damage. That can allow generation of well-defined ablated areas or microstructures especially at low laser fluence. In this section, results are presented that are related to single laser pulse effects and especially selective ablation of the first Ti nano-layer from the top of the experimental sample.

Apart of examination of morphology done by SEM and AFM, and composition by EDX, after single pulse laser radiation of the nano-layered thin film, we estimated damage/ablation threshold and Gaussian beam radius. In order to relate the morphological changes of partially ablated surface to temperature evolution on the surface and beneath it, we presented calculation of the temperature evolution.

*SEM and EDX analyses*

Firstly, we performed inspection of all single laser pulse spots by optical microscopy and by SEM at low magnification. After that, morphological and compositional changes of spots important for selective ablation were studied by SEM (Secondary and Backscattered Electron Imaging) at high magnifications and EDX, respectively. Considering the aim this study, we focused on single pulse irradiation with pulse energies of 0.5 µJ and higher. Single pulses with pulse energies lower than that did not ablate the upper Ti layer from the nano-layered structure.

Morphological changes induced by single fs laser pulses (at various energies) are evident in the SEM micrographs (Secondary Electron Imaging), shown in the Figure 1. Circular ablated region with sharp boundary is shown in all spots, however, the energy of the irradiation laser pulse determines whether there is a single or second step ablation. This distinction is noted in Fig.1 in which the letters I, II, III, IV stand for a non-irradiated surface, a single step ablation (characteristic for entire spot or for the region close to edges), the second step ablation (flat central or ring areas) and non-uniform center of the ablated region, respectively. A SEM analysis revealed that after irradiation with pulse energy of 0.5 µJ, the spot center remained flat with the clear edge of the ablated area (Fig.1.a1 and a2). The edge between ablated and non-ablated area was very sharp, indicating the exfoliation of the top layer. At pulse energy of 1 µJ the similar one step ablation was registered and the spot morphology was preserved, while the size of spot was expanded (not presented here). On the other hand, irradiation with higher pulse energy as 2 µJ caused two-step ablation (Fig. 1.b1 and b2). The spot consisted of an outer single step ablated region, and inner central region, second step ablation. Further increase in pulse energy caused a two-step ablation but with a new specific morphology (Fig. 1.c1 and c2).

SEM micrographs in BSE mode (Back-scattered Electron Images) of the Ti(5x(Al/Ti))/Si after single fs laser irradiation with different pulse energies- 0.5, 2, 5 and 10 µJ - are presented in Figure 2.(a-d). The BSE images help to obtain compositional maps of the sample but they provide only qualitative information about the surface composition. Thus, *brighter* areas are related to a greater average atomic number $Z$ in the sample, while lighter areas corresponds to a lower average $Z$. BSE images are limited to the polished samples, therefore, as information from the rest parts of the spots are not conclusive.

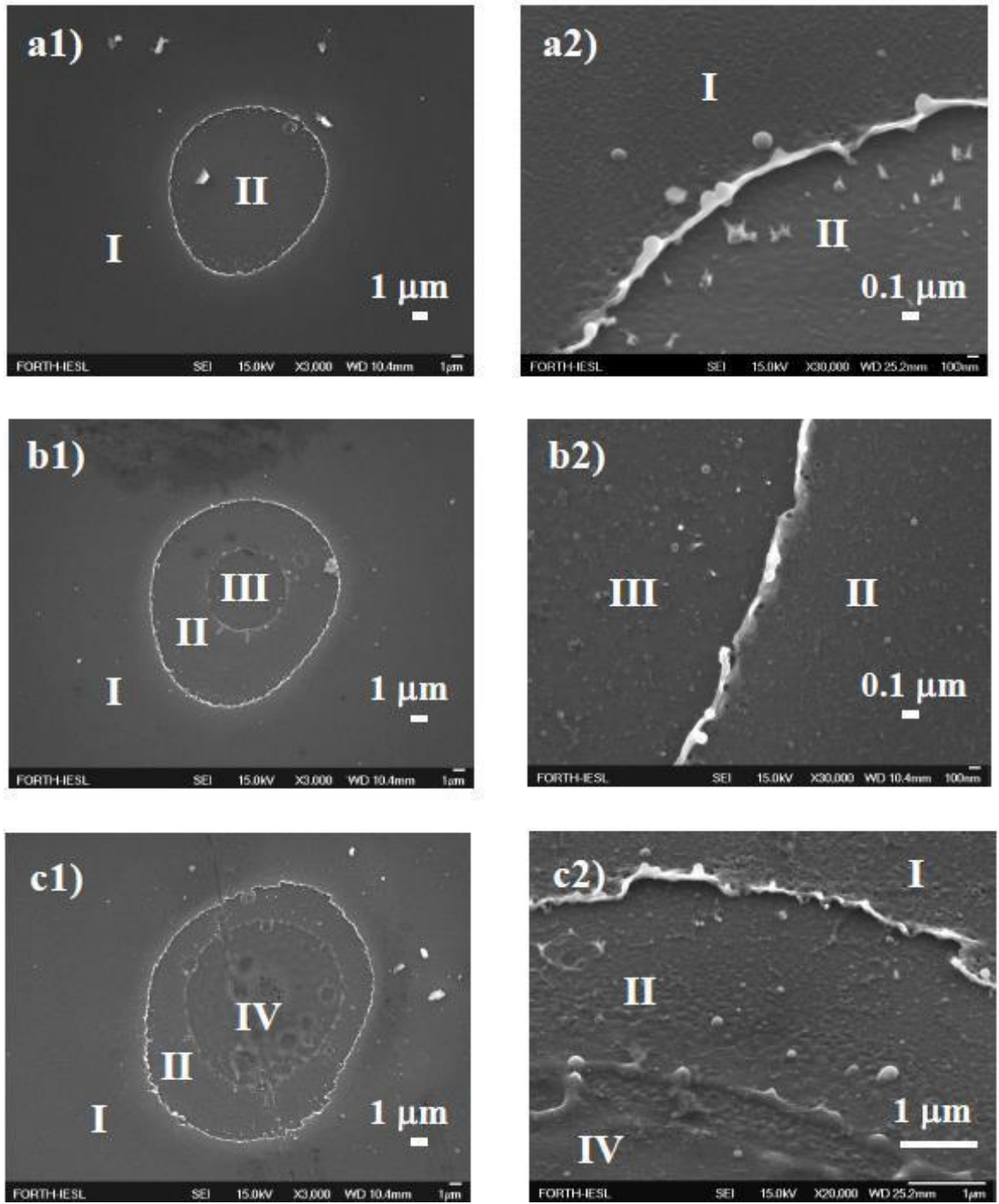

Figure 1. SEM images of Ti(5x(Al/Ti))/Si after single 170 fs laser pulse irradiation at different energies: Pulse energies: a1) and a2), 0.5 µJ; b1) and b2), 2 µJ; c1) and c2), 5 µJ.

It is evident that at pulse energy 0.5 µJ non-irradiated area I is brighter – rich in Ti - while the ablated area II is darker - reduced Ti (Fig.2.a). At pulse energy equal to 2 µJ, the area of single step ablation area II (Fig.2.b), shows the same difference in composition compering to area I. A similar difference of the non-irradiated areas and single step ablation regions are visible in Figures 2. c and d. Notably, the non-flat central parts of produced spots in the BSE micrographs are not accurate for composition mapping.

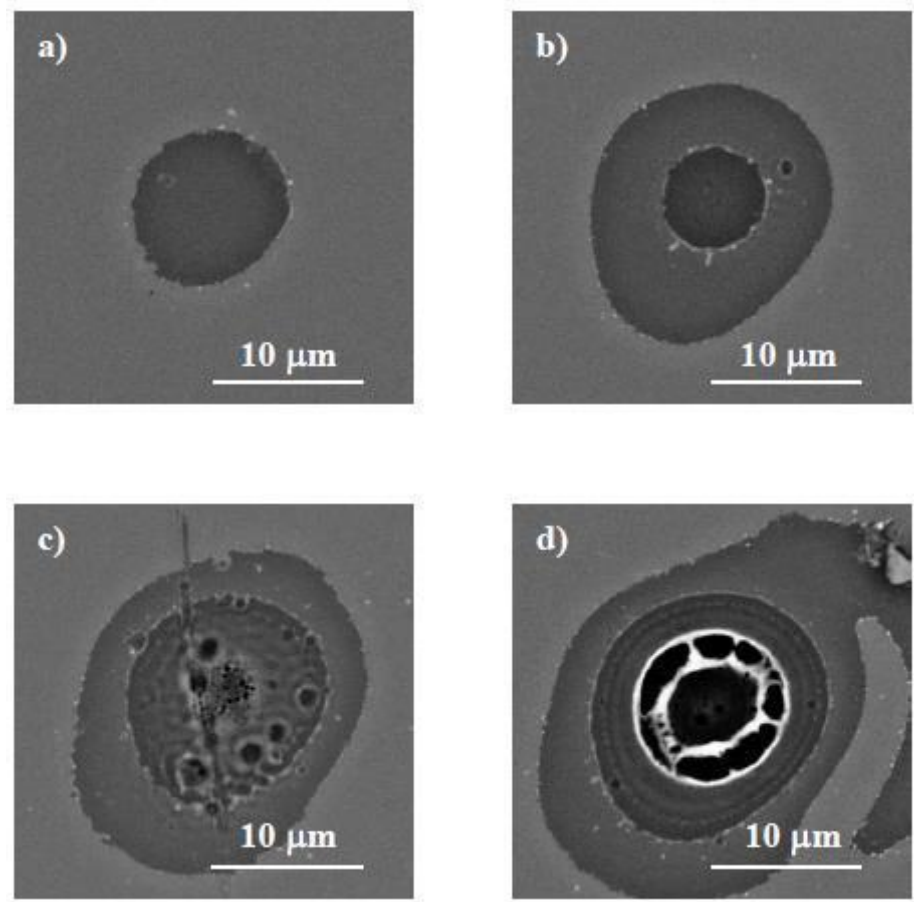

Figure 2. SEM micrographs (BSE images) after a single 170 fs laser pulse irradiation of Ti(5x(Al/Ti))/Si with pulse energy: a) 0.5 µJ, b) 2 µJ, c) 5 µJ and d) 10 µJ.

Quantitative analysis - EDX of the sample was performed for studying the difference in the elemental composition of non-irradiated thin film and ablated regions. Composition in weight percentage (wt%) of the regions I, II, III and IV marked on the Figures 1. is given in the Table 1. Composition of as deposited TF can be seen from values given in the first row of Table 1, denoted with I (average). Because of small TF thickness, signal from Si substrate is high. Concentration of Ti and Al as constitutive elements of as deposited TF/in the non-irradiated areas (regions I) were 11.00 and 3.98 (wt%), respectively.

| Positions from areas on Fig. 1 | O | Al | Si | Ti |
|---|---|---|---|---|
| I (average) | 1.33 | 3.98 | 83.69 | 11.00 |
| II a1 (0.5 µJ) | 1.11 | 3.55 | 87.15 | 8.19 |
| II b1 (2 µJ) | 0.91 | 3.43 | 87.59 | 8.07 |
| II c1 (5 µJ) | 1.30 | 3.60 | 86.66 | 8.43 |
| III b1 (2 µJ) | 1.06 | 2.71 | 89.97 | 6.25 |
| IV c1 (5 µJ) | 0.83 | 1.58 | 94.00 | 3.59 |

Table 1. The composition (wt%) of the surface of the sample from the regions marked with I, II, III and IV on the Figures 1 (a1, b1 and c1). The constituents of the sample are Al, Ti and Si, whereas oxygen originates from the surface contamination.

Depending on the laser pulse energy and ablated region on the spots, the changes in Al and Ti, as well as in Si concentrations were recorded. Thus in the regions II a1 and II b1, where selective ablation of Ti from the rest of the TF was happened, following concentrations were registered 8.19 (wt%) of Ti and 3.55 (wt%) of Al and 8.07 (wt%) of Ti and 3.43 (wt%) of Al, respectively. The Al concentration was close to the value of as deposited TF, while concentration of Ti was lower. Similar result was registered after irradiation with laser pulse energy of 5 µJ in region II c1 Based on the reduction of Ti concentration for approximately 3 %, this could correspond to a loss of upper Ti nano-layer from the thin film surface. The compositional analysis indicates that single step ablation was happened in this region. A small difference in Ti and Al concentrations can be due to the fact that at higher pulse energy the ablated region originates from used Gaussian beam. In the regions III and IV (III b1 pulse energy 2 µJ; IV c1 pulse energy 5 µJ), due to intensive ablation, the reduction of both Ti and Al concentrations were registered. The irradiation at the highest pulse energy (5 µJ) caused the smallest percentage of TF constituents with dominant presence of Si.

### *AFM analysis*

Apart from the morphology and compositional investigation, the depth of ablated areas was quantified by AFM analysis. The distance from the bottom of the ablated area to the surface (step height) of the nano-layer was measured on the laser spots after irradiation at pulse energies of 0.5, 1 and 2 µJ. At higher laser pulse energies AFM analyze was difficult because of very rough surface in the spots center. As an illustration of AFM analyze, the case of irradiation with pulse energy of 0.5 μJ is presented on the Figure 3.

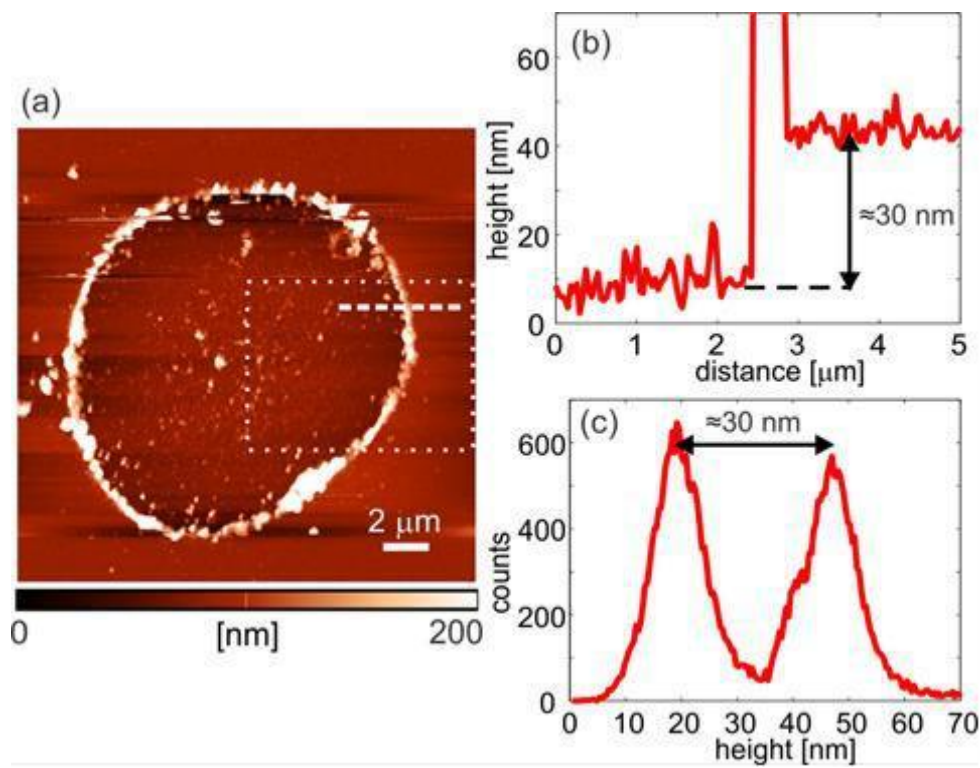

Figure 3. (a) topography of the Ti(5x(Al/Ti))/Si surface after single 170 fs laser pulse of pulse energy 0.5 μJ, (b) the cross section of the topography along the dashed line in part (a), and (c) the histogram of the rectangular region denoted in part (a) with dotted line. Both cross section and histogram show the step height of around 30 nm. Height is saturated in part (b) due to better visibility.

Fig. 3.a and b, as well as SEM micrographs (Fig. 1a and b, and Fig.2.a and b) have shown that partial/selective ablation of the Ti from Ti(5x(Al/Ti)) thin film was happened. The value of about 30 nm-the ablation step-is near the value of 27 nm that is thickness of upper Ti layer. This result confirmed that partial/selective ablation was occurred. Further AFM analysis of the ablation steps obtained after irradiation with 2 μJ revealed that the first step between regions I and II (Fig.1b1) was about 30 nm. The profile along II and III regions (Fig. 1b2) gave step value of 29.8 nm which could resemble to additional partial ablation of an Al layer and uncomplete ablation of the next Ti layer (Fig. 4). The results are consistent with changes in the concentration of Ti and Al species (EDX results), after irradiation with corresponding pulse energies.

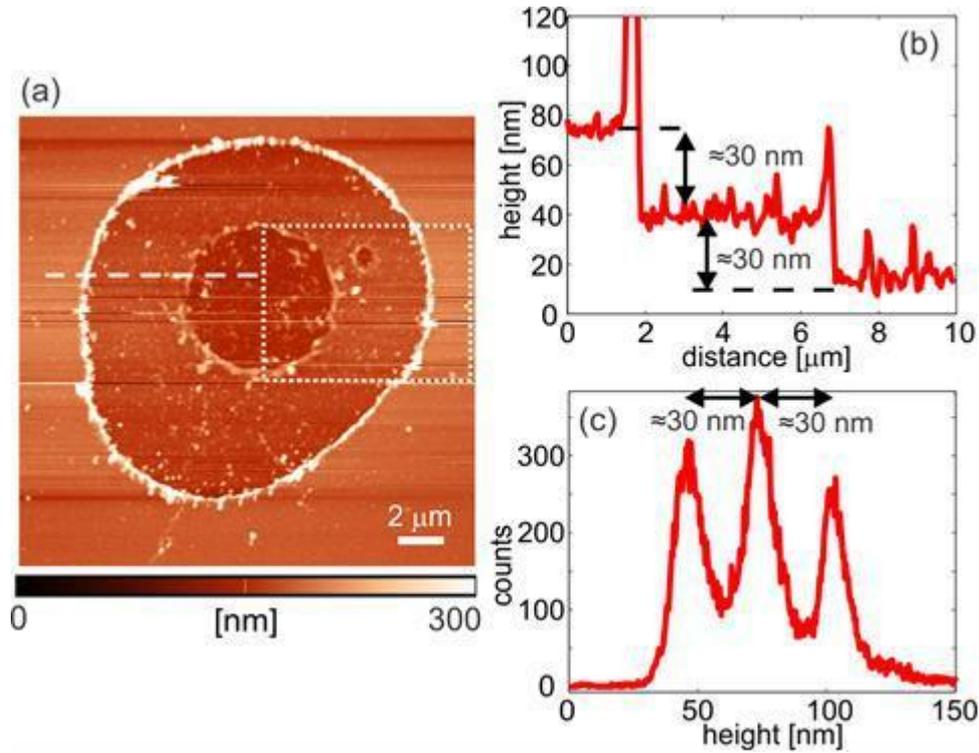

Figure 4. (a) topography of the Ti(5x(Al/Ti))/Si surface after single 170 fs laser pulse of pulse energy 2 µJ, (b) the cross section of the topography along the dashed line in part (a), (c) the histogram of the rectangular region denoted in part (a) with dotted line. Both cross section and histogram show two steps with a height of around 30 nm. Heigth is saturated in part (b) due to better visibility.

## *Ablation threshold*

The laser beam intensity at which damage/ablation occurs is known as laser induced damage threshold. In the case of ultra-short laser pulses, laser induced damage/ablation threshold can be precisely estimated. The threshold is a function of various parameters including laser wavelength, pulse duration, temporal and spatial profile, spot size and sample characteristics. Therefore, the value of reported damage/ablation fluence can considerably vary. Based on available published results, relating to damage/ablation threshold of bulk Ti (that is deposited as first layer in our experimental sample) for ultra-short laser pulse radiation, it was found a few data of threshold fluence: (i) Mannion, at all[23] in the case of 775 nm/150 fs radiation reported ablation threshold of $F_{th}$ = 280 mJcm$^{-2}$, (ii) Vorobyev and Guo[24] in the case of 800 nm/65 fs reported damage threshold of $F_{th}$ = 67 mJcm$^{-2}$, (iii) Hashida, at all[25], in the case of 800 nm/45 fs found ablation threshold of $F_{th}$ = 74 mJcm$^{-2}$ (iv) in agreement with theoretical prediction of Gamaly[10] ablation threshold values for typical metals and ultra-short laser radiation should be in the interval of 100-200 mJcm$^{-2}$.

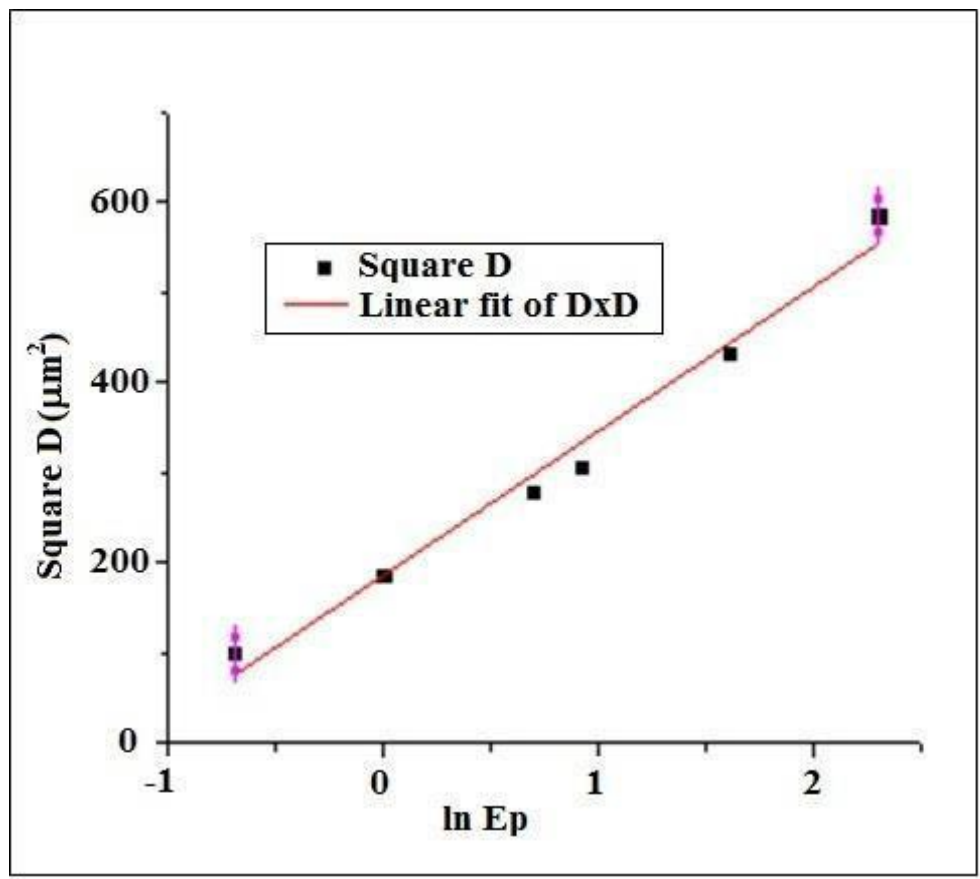

Figure 5. Ablation threshold fluence of $F_{th}$ = 250 mJcm$^{-2}$ was found from the squared diameter, $D^2$, of the ablated areas in Ti(5x(Al/Ti)) thin film, as a function of pulse energy Ln($E_p$) (pulse energy $Ep$ in µJ).

We measured ablation threshold and the beam radius following well known procedure[26,27]. Using the procedure, the Gaussian beam radius $w_o$ and ablation threshold fluence can be obtained using the diameters of the ablated areas $D$ versus the applied pulse energies $E_p$. Due to the linearity between energy and peak pulse laser fluence, one can determine $w_o$ by a linear least squares fit in the representation of $D^2$ as a function of ln($Ep$). The diameters D, recorded at different pulse energy, were taken from SEM micrographs of ablated regions. As a result, ablation threshold fluence $F_{th}$ = 250 mJcm$^{-2}$ and beam radius $w_0$ = 8.96 µm were estimated (Figure 5). The corresponding pulse energy to the threshold fluence is 0.31µJ. The beam radius value is close to the experimental one.

The experimental results showed that irradiation by single laser pulse with energy/fluence of 0.2/127(μJ/ mJcm$^{-2}$), that was below the value of the estimated ablation threshold, did not cause any visible changes of the surface, whereas the single pulse with energy/fluence of 0.5/320 (μJ/ mJcm$^{-2}$) was able to ablate the first Ti layer from the nano-layer thin film. In order to confirm this result we compare this finding with theoretical prediction concerning to temperature evolution for the fluence $F = 320$ mJcm$^{-2}$.

## *Theoretical Model-Simulation procedure*

To interpret the experimental observations, simulations have been performed to explore the thermal response of the multiple layered structure (Ti(5x(Al/Ti))) after irradiation with a single laser pulse of fluence $F = 320$ mJ/cm$^2$ and pulse duration $\tau_p$=170 fs. For the sake of simplicity, a much larger spot irradiation radius was considered compared to the thickness of the structure that allows a 1D analysis. The Two Temperature Model (TTM)[28] is used to describe the relaxation process following electron excitation due to laser heating (at time $t$ and at distance $z$ from the surface)

$$C_e^{(i)} \frac{\partial T_e^{(i)}}{\partial t} = k_e^{(i)} \frac{\partial^2 T_e^{(i)}}{\partial z^2} - G_{eL}^{(i)}\left(T_e^{(i)} - T_L^{(i)}\right) + S^{(i)}(z,t) \quad [S^{(i)} \equiv 0, \text{ for } i>1]$$

$$C_L^{(i)} \frac{\partial T_L^{(i)}}{\partial t} = k_L^{(i)} \frac{\partial^2 T_L^{(i)}}{\partial z^2} + G_{eL}^{(i)}\left(T_e^{(i)} - T_L^{(i)}\right) \tag{1}$$

$$S^{(1)}(z,t) = \frac{\alpha(1-R-T)\sqrt{4\log 2}\, F}{\sqrt{\pi}\tau_p} \exp\left(-4\log 2 \left(\frac{t-3\tau_p}{\tau_p}\right)^2\right) \exp(-\alpha z) \tag{2}$$

where $T_e$ ($T_L$) stand for the electron (lattice) temperature of layer $i$ ($i$=1,3,5,7,9,11 for Ti layer, $i$=2,4,6,8,10 for Al layer). It is also noted that the multiple-layered structure is placed on a Silicon substrate (in which $T_L^{(Si)}$ corresponds to its lattice temperature). The thermophysical properties of the materials such as electron and lattice heat capacity, ($C_e^{(i)}$, $C_L^{(i)}$), electron and lattice heat conductivity ($k_e^{(i)} \equiv k_{e0}^{(i)}\left(B^{(i)}T_L^{(i)} / \left(A^{(i)}\left(T_e^{(i)}\right)^2 + B^{(i)}T_L^{(i)}\right)\right)$[29], $k_L^{(i)} \sim .01 k_e^{(i)}$), electron-phonon coupling strengths ($G_{eL}^{(i)}$) and model parameters used in the simulations are listed in Table 2. While Eq.2 provides the general expression of the form of the source term due to material heating with a pulsed laser that includes the absorption coefficient $\alpha$, the reflectivity $R$ and the transmission coefficient $T$ of the material, the Transfer Matrix Method[30] is used to compute the optical properties of the top layer (Ti) after irradiation with pulsed laser of 1026 nm by taking into account the presence of the rest of the thin layers. Calculations yield $\alpha=4.89\times10^5 cm^{-1}$ [31], $T=0.0238$, $R=0.6939$, that indicate that ~31% of the energy will be absorbed in the first layer, while the transmitted part of the laser energy into the second layer (Al) is very small and it is not sufficiently high to excite the electrons in the rest of the layers (especially the second layer) and produce meaningful results. This argument justifies the use of a source term to describe laser heating of the first layer only while it is assumed laser energy is not transmitted into the next layers.

Eqs.1, 2 are solved using a finite difference scheme by using an iterative Crank-Nicolson scheme. As initial conditions, we choose thermal equilibrium at $T_e(z,t=0)=T_L(z,t=0)=300K$. Adiabatic boundary conditions are assumed on the surface (at $z=0$, $k_e^{(Ti)}\frac{\partial T_e^{(Ti)}}{\partial z}=k_L^{(Ti)}\frac{\partial T_L^{(Ti)}}{\partial z}=0$) and the back of the complex structure ($k_L^{(Si)}\frac{\partial T_L^{(Si)}}{\partial z}=0$). Furthermore, at the interface between two layers, the following conditions are applied: $T_L^{(Ti)}=T_L^{(Al)}$, $T_e^{(Ti)}=T_e^{(Al)}$, $k_L^{(Ti)}\frac{\partial T_L^{(Ti)}}{\partial z}=k_L^{(Al)}\frac{\partial T_L^{(Al)}}{\partial z}$, $k_e^{(Ti)}\frac{\partial T_e^{(Ti)}}{\partial z}=k_e^{(Al)}\frac{\partial T_e^{(Al)}}{\partial z}$ while on the interface between the last layer (Al) and the substrate (Si), the following boundary conditions are used: $T_L^{(Si)}=T_L^{(Al)}$ and $k_L^{(Si)}\frac{\partial T_L^{(Si)}}{\partial z}=k_L^{(Al)}\frac{\partial T_L^{(Al)}}{\partial z}$. Theoretical calculations of the electron and lattice temperatures based on the scheme described above yield a spatio-temporal evolution that is illustrated in Fig.6a. It is evident that the lattice temperatures attained in the first layer of Ti are large enough to induce evaporation of the material. By contrast, a very thin layer will be removed in the second layer (Al) while material with temperatures higher than the melting point, $T_{melting}$, are expected through fluid transport to determine the final surface profile on Al.

Although, the aim of the current work focuses on the description of the thermal response of the system, the fact that induced temperatures lead to phase transition or evaporation suggests that appropriate phase changes-related corrections need to be incorporated into the model. Thus, a thorough approach requires the inclusion of Navier-Stokes equations (to describe fluid dynamics) and relevant equations to account for evaporation[32,33,34]. Nevertheless, as the predominant aim of the work is to demonstrate that the laser energy used in the experiment is sufficient to remove the upper layer, simulations are performed, to first approximation, by ignoring hydrodynamics-generated effects. Furthermore, to simulate the mass removal mechanism, it is assumed that all lattice points that attain temperatures (derived from Eqs.1) larger than the boiling temperature $T_{boiling}$ are removed. The expected mass removal process of the first layer is illustrated in Fig.6a (*white* region indicates the ablated part of Ti that is not modelled at later times). On the other hand, the spatial distribution of the lattice temperature at $t$=10ps (Fig.6b) shows that, substantial temperature rise occurs only in the first three layers. It has to be emphasised that the choice of $T_{boiling}$ as a criterion to

|  | Ti | Al |
|---|---|---|
| **Parameter** | Value | Value |
| $G_{eL}$ [Wm$^{-3}$K$^{-1}$] | Fitting[5,35] | Fitting[35] |
| $C_e$ [Jm$^{-3}$K$^{-1}$] | Fitting[5,35] | Fitting[35] |
| $C_L$ [Jm$^{-3}$K$^{-2}$] | 2.3591×10$^6$ [36] | 2.42×10$^6$ [36] |
| $k_{e0}$ [Jm$^{-1}$s$^{-1}$K$^{-1}$] | 21.9 [36] | 235 [36] |
| $T_{melting}$ [K] | 1941 [36] | 934 [36] |
| $T_{boiling}$ [K] | 3560 [36] | 2743 [36] |
| $T_{critical}$ [K] | 15500 [37] | 8550 [38] |
| $A$ [s$^{-1}$K$^{-2}$] | Fitting[5,35] | Fitting[35] |
| $B$ [s$^{-1}$K$^{-1}$] | Fitting[5,35] | Fitting[35] |

Table 2. Simulation parameters chosen for Ti and Al

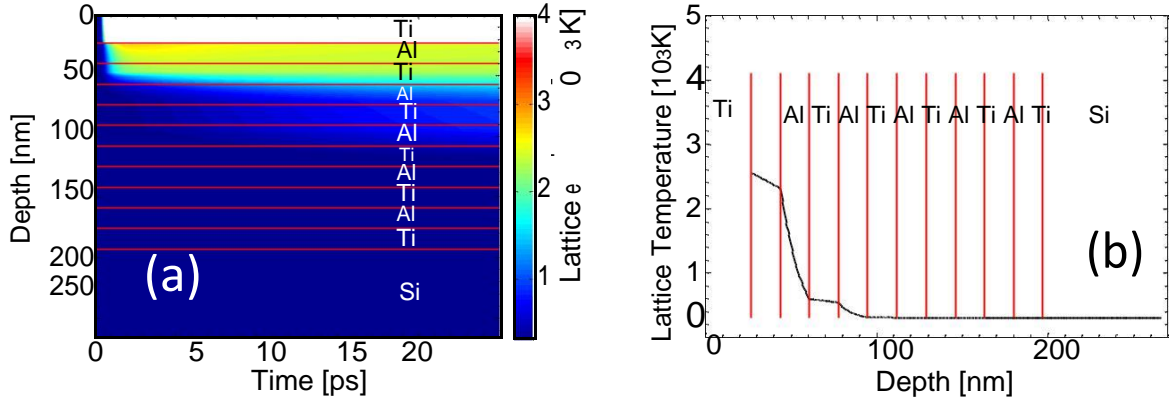

Figure 6. (a) Lattice temperature field evolution in depth, perpendicular to the surface of the sample (*white* region indicates temperatures leading to material removal; *horizontal* dashed lines indicate the size of each layer). (b) Spatial lattice temperature profile at $t$ = 10 ps, *vertical* dashed lines indicate the size of each layer). ($F$ = 320 mJ/cm$^2$, $\tau_p$ = 170 fs, Laser beam wavelength is 1026 nm) determine whether ablation occurs differs from the threshold value used in other studies (~0.90$T_{critical}$ where $T_{critical}$ is the critical point temperature[33,39,40].

Our simulations show that, for Ti, that threshold is extremely high (~13950K[37]) while at fluence $F$=320 mJ/cm$^2$ the maximum temperature attained from the material is comparable to $T_{boiling}$. Therefore, as simulation results predict that the lattice temperatures in the upper Ti layer are of the order of $T_{boiling}$ and experimental observations confirm that the Ti layer is entirely removed, the boiling temperature is considered as a reliable mass removal threshold choice.

## 4. Conclusion

In this paper, we presented results following irradiation of Ti/Al nano-layered thin film with ultra-short laser pulses. Single pulse irradiation was conducted in the air with focused and linearly polarized femtosecond laser beam- of 1026 nm wavelength and pulse duration of 170 fs. Applied laser pulse energy/fluence was from 0.2/127(μJ/mJcm$^{-2}$) to 10/6.37(μJ/Jcm$^{-2}$). Morphological and compositional changes of the sample surface were monitored by scanning electron microscopy (SEM - SE and BSE imaging), energy-dispersive X-ray spectroscopy (EDX) and atomic force microscopy (AFM). The smallest laser beam energy/fluence was not able to ablate nano-layer surface; however the highest one caused complete ablation with significant change of Si substrate. The ablation threshold, for used laser parameters and sample characteristics, was found $F_{th}$ = 250 mJcm$^{-2}$ or 0.31 μJ. We concluded that irradiation of the sample with single laser pulse of 0.5/320($\mu$J/mJcm$^{-2}$) caused selective ablation of the Ti from the nano layer thin film.

Apart from experimental examination of morphology and compositional changes, the theoretical calculation of the surface temperature was performed. Our simulations show that for Ti at fluence $F = 320$ mJ/cm$^2$, the maximum attained temperature in the material is comparable to $T_{boiling}$. Therefore, simulation results predict that the lattice temperatures in the upper Ti layer are of the order of $T_{boiling}$ and experimental observations confirm that the Ti layer is entirely removed. It is noted that the boiling temperature is considered as a reliable mass removal threshold choice.

In conclusion, it was shown that single fs laser pulse, at certain pulse energy, is able to remove the first nano-layer from the reactive multiple layer thin film without noticeable changes of the sample. This partial ablation may provide an additional route for controlling and optimizing the outcome of laser nano/micro-processing of RMTF.

**Acknowledgements**

B. Gakovic, S. Petrovic and B. Vasic acknowledge financial supports from the Ministry of Education, Science and Technology of Serbia (Project No.III-45016 and OI-171005), and European COST association (COST Action MP.1208). G.D.T, E.S and E.S acknowledge financial support from the project *LiNaBioFluid*, funded by the European Union's H2020 framework programme for research and innovation under Grant Agreement No. 665337 and the project *Nanoscience Foundries and Fine Analysis* (NFFA)–Europe H2020-INFRAIA-2014-2015 (Grant agreement No 654360).